\documentclass[10pt,conference]{IEEEtran}
\IEEEoverridecommandlockouts
\usepackage{cite}
\usepackage{amsmath,amssymb,amsfonts}
\usepackage{algorithmic}
\usepackage{graphicx}
\usepackage{textcomp}
\usepackage{xcolor}
\usepackage{makecell}
\usepackage[linesnumbered,boxed]{algorithm2e}
\usepackage{multirow}
\usepackage{multicol}
\usepackage{hyperref}
\def\BibTeX{{\rm B\kern-.05em{\sc i\kern-.025em b}\kern-.08em
    T\kern-.1667em\lower.7ex\hbox{E}\kern-.125emX}}

\begin{document}

\title{Geofeed Adoption and Authentication
}

\author{\IEEEauthorblockN{Dipsy Desai, Kicho Yu, Sulyab Thottungal Valapu}
\IEEEauthorblockA{
\textit{University of Southern California},
Los Angeles, CA \\
\{deepakde, kichoyu, thottung\}@usc.edu
}

}

\maketitle

\begin{abstract}
IP Geofeed is a recently proposed informational standard that allows network operators to publish the geographical location of deployed IPv4 and IPv6 prefixes. In this work we study the adoption of IP geofeed, assess deployment of geofeed at Regional Internet Registry and Autonomous System levels, and analyze adherence to RFC 8805 and RFC 9092 in deployed geofeeds. We evaluate the authentication mechanism proposed in RFC 9092 and find that it lacks key features from a security perspective. We propose a novel approach to simplify the authentication of geofeeds and assess its efficiency using different benchmarks. Our findings highlight the challenges in current geofeed adoption and the potential for improving both security and scalability in geofeed validation processes.

\end{abstract}

\begin{IEEEkeywords}
IP Geofeed, Regional Internet Registry, Autonomous Systems, Authentication of Geofeed
\end{IEEEkeywords}

\section{Introduction}
IP geolocation is widely recognized for its ability to enhance user experience. For example, it enables selecting geographically closer servers to reduce latency, tailoring user experience based on location (like language or currency), and delivering context-aware search results (such as ``events this weekend'').
Current state-of-the-art solutions employ a variety of techniques, including multilateration based on latency measurements~\cite{gueye_constraint-based_2006} to estimate the approximate geolocation of hosts, and more recent advancements like machine learning models \cite{Ding2023a_GNN_geolocation} or search engine clicks \cite{Dan2022a_gelocation_clicks} to improve IP geolocation. 

Despite these advancements, IP geolocation alone is not sufficient to track real-time changes in IP address assignments, highlighting the need for a mechanism to proactively signal such changes. For example, these methods can become temporarily \textit{stale} due to changes in prefix deployments, like Internet Service Provider (ISP) renumbering. To address this, the IETF introduced \emph{geolocation feeds}, or \emph{geofeed}, a technique enabling network operators to publish the geolocation of deployed IP prefixes. The standard is split across two RFCs: RFC 8805~\cite{kline_format_2020} defines the format of geofeed files, while RFC 9092~\cite{bush_finding_2021} details the standard methods for publishing and utilizing geofeed files.

Given its status as a nascent technology in current Internet standards, there is limited information on the adoption of geofeed by network operators. In this work, we perform an initial measurement study of geofeed adoption at multiple levels of Internet organization, specifically the Regional Internet Registry (RIR) and Autonomous System (ASes). We analyze the extent to which geofeeds adhere to RFC standards and explore the shortcomings of the \emph{authentication} procedures described in RFC 9092. Based on our analysis, we propose a novel, secure, and scalable two-step authentication method for geofeed publication and validation.

\section{Analysing Geofeed Adoption} \label{sec:analysis}
RFC 9092~\cite{bush_finding_2021} specifies that geofeed information should be included in the \texttt{inetnum}, \texttt{inet6num}, or \texttt{NetRange} database classes as defined by the Routing Policy Specification Language (RPSL)~\cite{murphy_routing_1999}. To gather geofeed data, our first step was to collect the relevant \texttt{inetnum}, \texttt{inet6num}, or \texttt{NetRange} records from all five RIRs, namely AFRINIC, APNIC, ARIN, LACNIC and RIPE NCC, during February and March 2024. These records were queried from publicly available RIR databases.

Once the records were obtained, we parsed them to extract the URLs of the \texttt{csv} files containing the actual geofeed data. These geofeed URLs were then used to download the CSV files associated with each RIR's data. During this phase, we encountered some challenges, for example, approximately 7.76\% of the geofeed URLs were inaccessible due to DNS resolution failures, connection timeouts, or various HTTP errors. Specifically, out of the total 1547 URLs queries, 1427 were accessible, while the remaining URLs failed to establish connections  (the most common issue), or resulted in HTTP 404 (Not Found) errors.

After successfully gathering the data, we proceeded with a detailed analysis aimed at answering several key research questions related to geofeed adoption, compliance with RFC standards, and the efficacy of geofeed authentication methods.

\subsection{Do certain RIRs adopt geofeed more quickly than others?} \label{sec:analysis/rir}
To better understand the deployment of geofeed, we first look at its adoption at the RIR level.
Table~\ref{tab:rirstats} shows the per-RIR breakdown of \texttt{inetnum}s and \texttt{inet6num}s having geofeed information.
Noticeably, RIPE NCC leads significantly in the count of both \texttt{inetnum}s and \texttt{inet6num}s with geofeed entries among other RIRs, as well as the number of ASes with geofeed entries.
In fact, RIPE accounts for 82.04\% and 88.24\% of geofeed-enabled \texttt{inetnum}s and \texttt{inet6num}s respectively.
However, overall, only 0.25\% of \texttt{inetnum}s and 0.30\% of \texttt{inet6num}s have associated geofeed entries.
This indicates that geofeed adoption is still in its early stages.

\begin{table}[htb]
  \begin{tabular}{|l|l|l|l|l|l|}
    \hline
    \multirow{2}{*}{\textbf{RIR}} & \multicolumn{2}{c|}{\textbf{\texttt{inetnum}}} & \multicolumn{2}{c|}{\textbf{\texttt{inet6num}}} & \multirow{2}{*}{\textbf{\# AS}}                                   \\ \cline{2-5}
                                  & Count                                          & Fraction                                        & Count                           & Fraction        &               \\ \hline
    AFRINIC                       & 421                                            & 0.28\%                                          & 24                              & 0.07\%          & 19            \\ \hline
    APNIC                         & 871                                            & 0.07\%                                          & 141                             & 0.14\%          & 156           \\ \hline
    ARIN                          & 1375                                           & 1.84\%                                          & 206                             & 0.29\%          & 440           \\ \hline
    LACNIC                        & 58                                             & 0.01\%                                          & 16                              & 0.06\%          & 21            \\ \hline
    RIPE                          & 12447                                          & 0.30\%                                          & 2905                            & 0.34\%          & 1417          \\ \hline
    \textbf{Total}                & \textbf{15172}                                 & \textbf{0.25\%}                                 & \textbf{3292}                   & \textbf{0.30\%} & \textbf{1907} \\ \hline
  \end{tabular}
  \caption{
    Number and fraction of \texttt{inet[6]num}s per RIR with geofeed records, and number of ASes with at least one geofeed record.
    Note that the same AS may get counted under multiple RIRs based on available records.
  }
  \label{tab:rirstats}
\end{table}

\begin{figure}[htb]
  \centering
  \includegraphics[width=\linewidth,trim={1cm 2cm 1cm 3cm},clip]{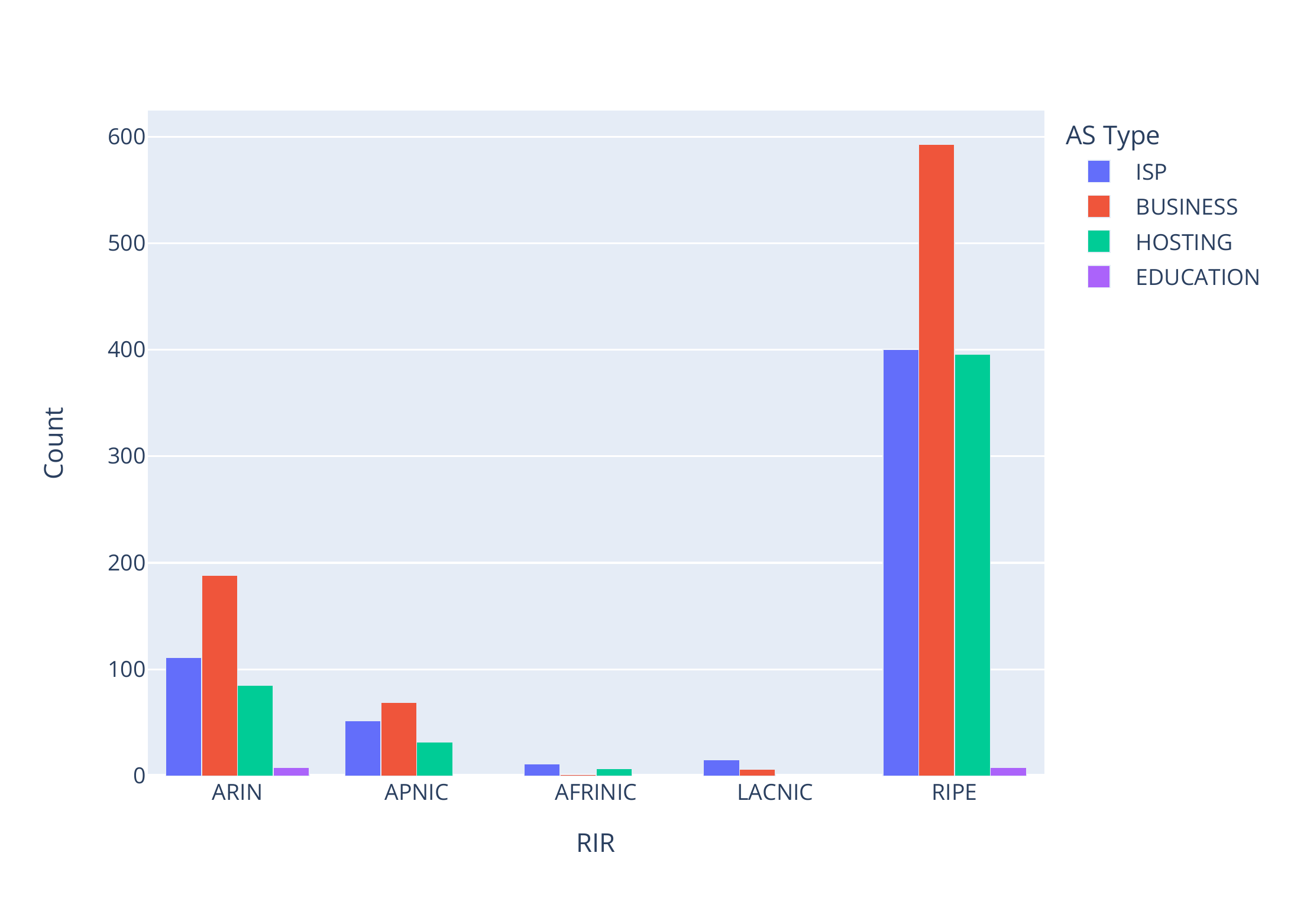}
  \caption{Category-wise breakdown of ASes that have geofeed records, grouped by RIR.}
  \label{fig:as_category_breakdown}
\end{figure}

\subsection{Do specific categories of ASes adopt geofeed more rapidly?} \label{sec:analysis/as}
Since the goal of geofeed is to aid in geolocating IP addresses, different categories of autonomous systems may have different levels of interest in its adoption.
For example, ISPs may be motivated to use geofeeds to signal changes in the geographical deployment of customer prefixes, whereas educational institutions may be less inclined to do so since their locations are usually fixed.
Hence, it is worth examining whether specific categories of ASes have been adopting geofeed more rapidly than others.

To answer this question, we used the AS information API provided by \texttt{ipinfo.io} to sort geofeed-enabled ASes within each RIR into ISP, Business, Hosting, and Education categories.
The results, presented in Figure~\ref{fig:as_category_breakdown} above, indicate that ASes belonging to the Business category lead geofeed adoption, followed by ISP and Hosting.

Our analysis primarily focuses on the RIR and AS level due to the data's higher-level structure. The records we gathered from the RIR databases include information about IP address ranges assigned to entire ASes or RIRs, but do not provide the level of detail found at the individual IP prefix level. As such, our analysis was constrained by the granularity of the available data. While this high-level analysis offers valuable insights into geofeed adoption and compliance, we recognize that examining geofeeds at a more granular level, such as individual IP prefixes (e.g., /32 for IPv4 or /64 for IPv6), could yield a more precise understanding of geofeed accuracy and adoption. Future work could explore this, if data at this finer granularity becomes available. For a visual representation of our findings, see Appendix \ref{sec:analysis/viz}, which highlights the global and some regional geofeed adoption patterns.

\subsection{Do geofeeds adhere to RFCs?} \label{sec:analysis/rfc}
RFCs 8805 and 9092 specify strict requirements for formatting and publishing geofeed.
These requirements are necessary to ensure the integrity of geofeeds as well as provide structural uniformity for parsers.
For our analysis, we selected a subset of the requirements specified by the RFCs that we identified to be ``important'' from the perspective of a consumer of geofeed data.
We now discuss the results of our RFC adherence analysis.

\subsubsection{RFC 9092}
RFC 9092 is concerned with publishing and discovering URLs of geofeed \texttt{csv} files.
We consider adherence of published geofeeds to the following specifications:

\begin{enumerate}
  \item Geofeeds in \texttt{inet[6]num} entry using the \texttt{remarks:} attribute must be formatted as follows (\texttt{Geofeed} is case sensitive):
  \texttt{remarks: Geofeed \\https://example.com/geofeed.csv}
  \item Geofeeds in \texttt{inet[6]num} entry using the \texttt{geofeed:} attribute must be formatted as follows: 
  \texttt{geofeed: https://example.com/geofeed.csv}
  \item Geofeed URLs must use \texttt{https}.
\end{enumerate}

The RFC also mentions that apart from using \texttt{remarks:}, geofeeds can be also published in a \texttt{geofeed:} attribute of the \texttt{inetnum} or \texttt{NetRange} object once RPSL supports that attribute, however, we have not found any in support of this.

\begin{figure}[htb]
  \centering
  \includegraphics[width=\linewidth,trim={0.75cm 2cm 0.7cm 3cm},clip]{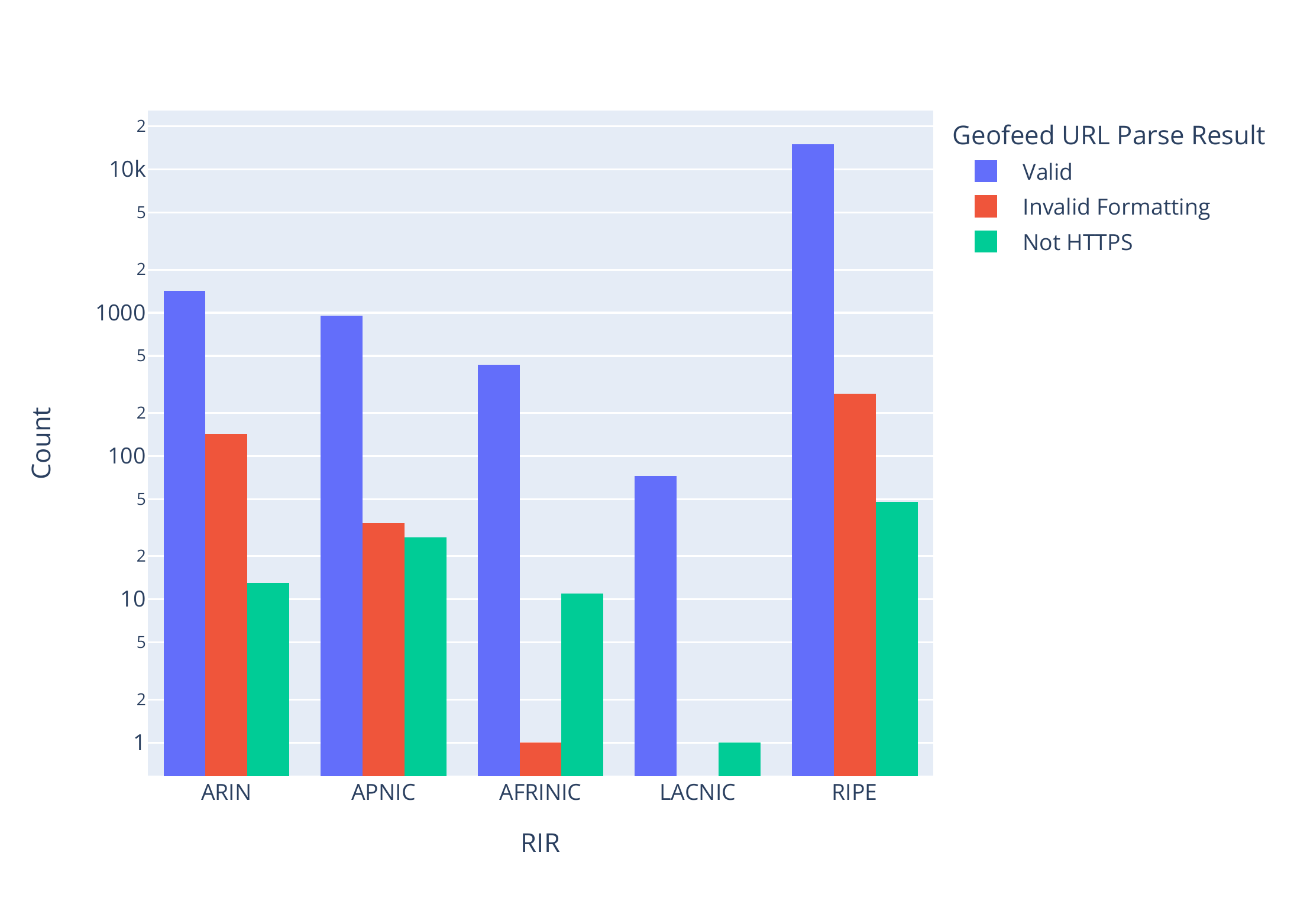}
  \caption{Results of RFC 9092 adherence analysis.}
  \label{fig:rfc_9092_adherence}
\end{figure}

If an \texttt{inet[6]num} passes all three checks, we categorize it as ``valid''.
Otherwise, we categorize it as ``invalid formatting'' or ``not HTTPS'' based on which check(s) it failed.
The analysis results are illustrated in Fig.~\ref{fig:rfc_9092_adherence}.
In particular, we note that 100 (0.54\%) \texttt{inet[6]num}s publish geofeed \texttt{csv}s over the unsecure \texttt{http} protocol.

\subsubsection{RFC 8805}
RFC 8805 is concerned with the content of the geofeed \texttt{csv} files.
We consider the following specifications for our analysis:

\begin{enumerate}
  \item Geofeed \texttt{csv} files must use UTF-8 character encoding and CRLF line breaks.
  \item All lines in a geofeed \texttt{csv} file must contain the following five fields in a comma-separated (no-spaces) format: \texttt{ip\_prefix,alpha2code,region,city,\\postal\_code}
  \item All fields except \texttt{ip\_prefix} can be empty, but the requisite number of commas must be present.
  \item The \texttt{ip\_prefix} field must be either a single IP address or an IP prefix in CIDR notation. The \texttt{alpha2code} and \texttt{region} fields, if non-empty, must be ISO country or region codes conforming to ISO 3166-1 alpha 2 and ISO 3166-2 respectively.
\end{enumerate}

\begin{figure}[htb]
  \centering
  \includegraphics[width=\linewidth,trim={0.75cm 2cm 0.75cm 3cm},clip]{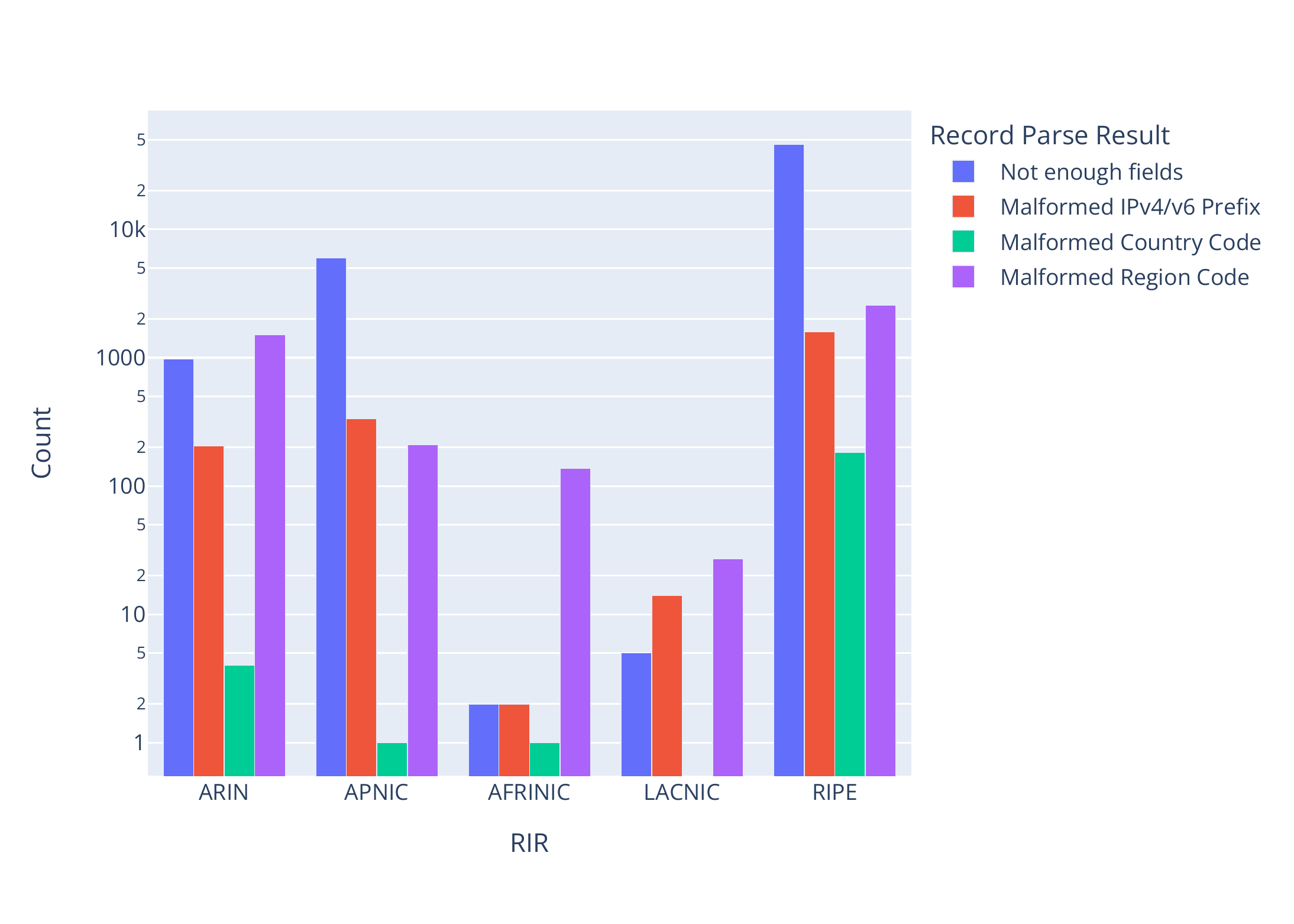}
  \caption{Results of RFC 8805 adherence analysis.}
  \label{fig:rfc_8805_adherence}
\end{figure}

If a \texttt{csv} line fails any of the check(s), we categorize it as ``malformed''.
We find that 511035 lines (89.51\%) out of 570909 are valid.
It is particularly alarming that 10.49\% of all geofeed lines are malformed and hence unusable.
For a closer look at the reasons for malformed lines, we further categorize malformed lines into ``not enough fields'', ``malformed IP prefix'', ``malformed country code'' and ``malformed region code'' based on which check(s) it failed.
The results, shown in Fig.~\ref{fig:rfc_8805_adherence}, indicate that not having enough fields, malformed IP prefix and malformed region codes are the major reasons for malformed lines.
We also find that, although all geofeed files use UTF-8 encoding, only 393 out of 1427 (27.54\%) use CRLF line breaks.

\section{Authenticating Geofeeds}
\label{sec:auth}
Geofeed data, which provides critical information about the geolocation of IP prefixes, must be accurate for consumers. RFC 8805 \cite{kline_format_2020} outlines factors for authenticating geofeed data, specifying procedures for validation before consuming self-published geofeed data. RFC 9092 \cite{bush_finding_2021} defines detailed authentication procedures, requiring the publisher to use Cryptographic Message Syntax (CMS) to create detached signature. The consumer must validate the signature, ensuring:

\begin{itemize}
    \item The signer's certificate is part of the current manifest and covered by the Resource Public Key Infrastructure (RPKI) certificate \cite{bush_finding_2021}.
    \item Path validation is carried out by using RPKI repositories (failure results in invalidation) --- \textit{Problem 1}
    \item The signed data is validated using the signer's public key certificate.
    \item The IP Address Delegation Certificate Extension \cite{rfc3779} covers all address ranges in the geofeed file (failure to match results in invalidation) --- \textit{Problem 2}
\end{itemize}

However, based on our findings, these requirements are too rigid due to:

\begin{itemize}
    \item Mismatches between the RPKI certificate resources and the current manifest.
    \item Limited compliance with certificate path validation, as only one AS adheres to this requirement mentioned in \textit{Problem 1}.
    \item Many ASes combine data into shared files, conflicting with the requirement that geofeed file must cover all ranges as mentioned in \textit{Problem 2}.
\end{itemize}

These shortcomings highlight the need for better integrity, ownership, and accuracy verification in geofeed data. To address this, we propose a new method to geofeed authentication, described in the next subsection.

\subsection{Two-Step Approach}

In this section, we describe our two-step approach to authenticate geofeed data. The first step involves authenticating the publisher of the geofeed, while the second step focuses on authenticating the geofeed data itself. By using this approach, we achieve the following objectives:
\begin{itemize}
    \item Publisher Authetication: We verify the geofeed information for only a subset of prefixes owned by an AS, while also enabling the verification of multiple publishers to a shared file.
    \item Data Integrity and Trustworthiness: We ensure data integrity, determine authoritativeness, and establish non-repudiation for the published geofeed data.
\end{itemize}

\subsubsection{\textbf{Authenticating the Publisher}}
To authenticate the publisher of geofeed data, we propose leveraging a Public Key Infrastructure (PKI). Unlike the approach in \cite{bush_finding_2021}, we rely on non-RPKI based repositories, enhancing the flexibility of the system for validation and verification. The PKI system binds the public key to the identity of the owner through the issuance of a certificate, ensuring the authenticity of the publisher. 

Consumers can verify the authenticity of a geofeed publisher by tracing its certificate to a trusted Certificate Authority (CA), like Verisign. Once the CA confirms the certificate’s validity, the consumer can trust that the publisher is who they claim to be, ensuring identity assurance.

\begin{figure}[htb]
  \centering
  \includegraphics[width=\linewidth,,trim={4cm 3.5cm 4cm 2cm},clip]{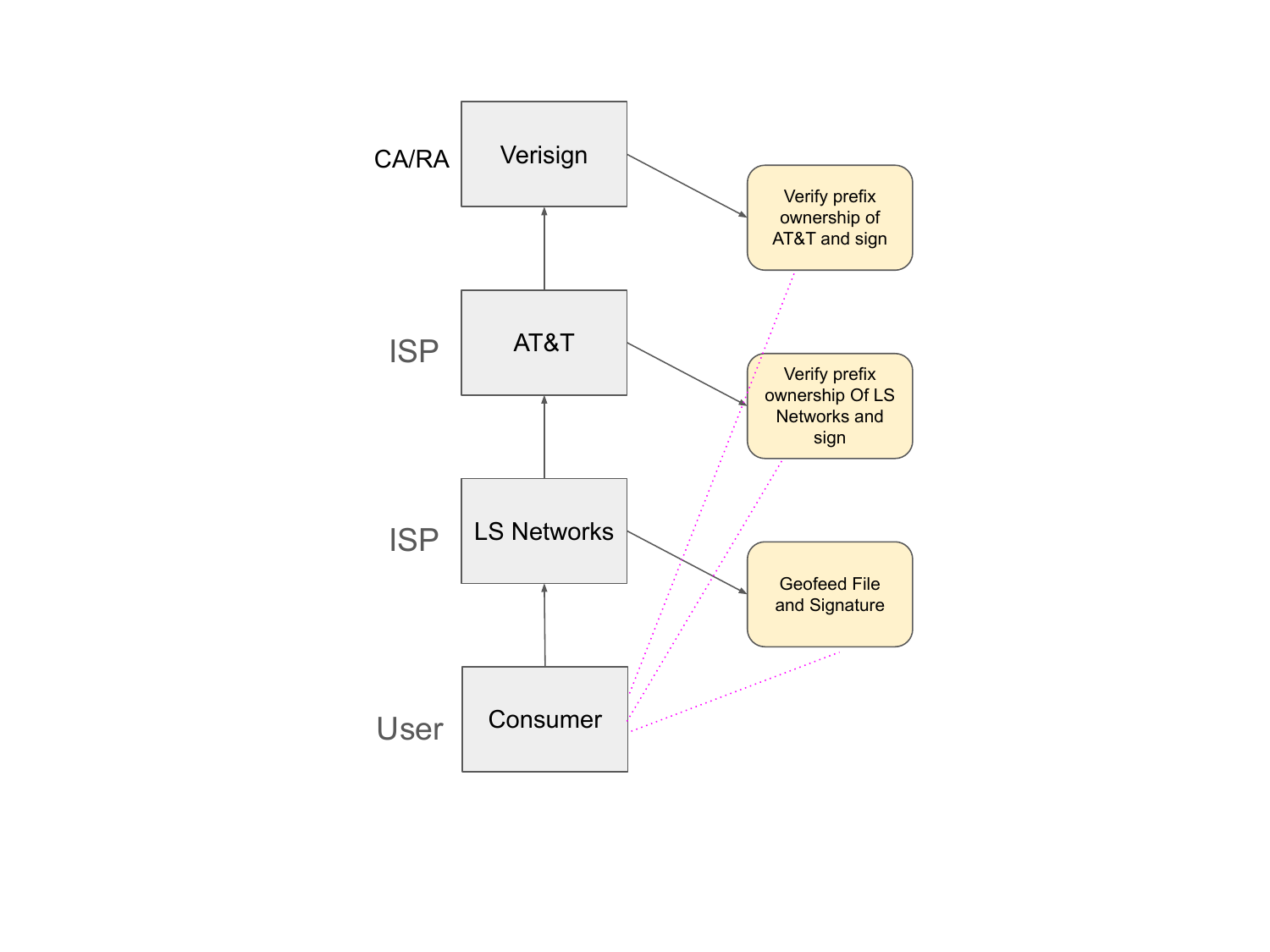}
  \caption{Authenticating Publisher and Geofeed Data through Digital Signatures}
  \label{fig:Authenticating_Geofeed_Data}
\end{figure}

In our experiments, we simulated the PKI ecosystem by generating over 1,800 unique certificates, each tied to an AS publishing geofeed data. For example, in Figure \ref{fig:Authenticating_Geofeed_Data}, we show three unqiue entities: a consumer, ISPs (ex: LS Networks, AT\&T), and a Certificate Authority / Registration Authority (CA/RA) (ex: Verisign). LS Networks publishes a geofeed, signs it with its private key, and attaches a certificate. The consumer can then verify the certificate using the CA’s public key, confirming the publisher’s identity. This method mirrors the successful use of PKI in web authentication and can be easily adapted to geofeed validation, providing a reliable means to verify the authenticity of geofeed data.

In Fig. \ref{fig:Authenticating_Geofeed_Data}, the solid line with an arrow represents a request to fetch a signed geofeed file from an entity higher in the hierarchy, such as a consumer requesting a geofeed data from LS Networks, and LS Networks requesting a signed geofeed file from AT\&T, and so on. The dotted magenta line indicates the consumer accessing signed geofeed data from various entities, such as ISPs or RIRs. This process ensures the integrity of data and validates the publisher’s identity, with each signature providing a layer of trustworthiness for the geolocation information.

\subsubsection{\textbf{Authenticating Geofeed Data}}
Geofeed data is generally stable, with IP prefix geolocations rarely changing. This stability supports iterative signatures, ensuring only authorized entities can publish geofeed information.

For example, in Fig. \ref{fig:Authenticating_Geofeed_Data}, LS Networks signs a geofeed file, and consumers can verify this signature using LS Networks' previously validated public key. If the consumer trusts LS Networks, they can use the geofeed data directly. If further verification is needed, a larger ISP like AT\&T can confirm LS Networks' authority over the prefixes, either through internal records or RPKI repositories. Once verified, AT\&T can sign the geofeed file or LS Networks' signature, and the consumer can verify it with AT\&T's public key. If the consumer trusts AT\&T, they can accept the geofeed data.

If doubts persist, the process can continue further up the chain. For instance, ARIN (an RIR) may hold a prefix like 120.0.0.0/8 and lease it to AT\&T, asserting AT\&T's ownership of the prefix. AT\&T, in turn, leases a sub-prefix (e.g., 120.1.1.0/24) to LS Networks, which can then validate LS Networks’ geofeed data for that specific prefix. A CA like Verisign can add another layer of validation by confirming AT\&T's prefix ownership and signing the data. This cascading verification process ensures that geofeed data is trusted at each level, with entities like ARIN asserting prefix ownership, and ISPs like AT\&T and LS Networks providing further validation.

\begin{table}[htb]
  \begin{tabular}{|l|l|l|l|l|}
    \hline
    \multirow{2}{*}{\textbf{\makecell{Prefix \\Comparison}}} & \multicolumn{2}{c|}{\textbf{\texttt{RIPE}}} & \multicolumn{2}{c|}{\textbf{\texttt{ARIN}}}                \\ \hline \cline{2-4} & \makecell{RPKI \\Repository} & ipinfo.io & \makecell{RPKI \\Repository} & ipinfo.io \\ \hline
    Correct/Match & 1109 & 6755 & 97 & 1104 \\ \hline
    Incorrect & 1175 & 221 & 134 & 209  \\ \hline
    Missing & 12244 & 7552  & 1294 & 212   \\ \hline
    \multirow{2}{*}\textbf{Total} & \multicolumn{2}{c|}{\textbf{14582}} &  \multicolumn{2}{c|}{\textbf{1525}} \\ \hline
  \end{tabular}
  \caption{Prefix ownership comparison for RIPE and ARIN based geofeed data.}
  \label{tab:prefixownershipcomparison}
\end{table}

\subsubsection{\textbf{Evaluation of proposed approach}}
To assess prefix ownership, we compared data from RIR databases with two secondary sources: RPKI repositories and ipinfo.io (see Table \ref{tab:prefixownershipcomparison} for RIPE and ARIN). We found low match rates in RPKI repositories (RIPE at 7.6\%, ARIN at 6.4\%), reflecting RPKI’s early adoption and limited coverage \cite{manrs_rpki_deployment}. However, match rates were much higher when compared to ipinfo.io (e.g., RIPE at 46.3\%, ARIN at 72.4\%). This suggests that RPKI has room for improvement in coverage and adoption, while external datasets like ipinfo.io can supplement RPKI for more effective prefix verification. Simulating the signing and validation process for over 1,800 unique certificates from ASes publishing geofeed data, we observed higher match rates with ipinfo.io, demonstrating the flexibility and comprehensiveness of our approach, particularly for leased or reassigned prefixes.

Whereas RPKI focuses on IP address ownership and ROAs (Route Origin Authorizations) for prefix validation, our method authenticates geofeed data through a PKI-based chain of trust, verifying both prefix ownership and the publisher's identity. Unlike RPKI’s emphasis on routing, our approach ensures the authenticity of geolocation data, which is crucial for decision-making. We eliminate the need for ROAs, as publishers directly sign geofeed data. This method scales well, since geofeed data changes infrequently, and requires only occasional updates to signatures. Validation involves checking certificates and signatures in the trust chain, providing flexibility compared to RPKI’s centralized prefix validation.

We acknowledge the study's limitation due to the lack of real-world data for testing our proposed authorization method, which, while simple and effective in theory, requires validation in practical settings.


\section{Conclusions and Future Work} \label{sec:conclusion}
In conclusion, our analysis of geofeed data shows that RIPE lead in adoption among RIRs, with ASes in the Business category driving geofeed use. However, publishers do not fully adhere to proposed standards, highlighting a gap between recommended practices and real-world implementation. To address these challenges and enhance the validation of geofeed data, we propose a multi-step approach for authentication.

Future work includes real world testing to assess performance and exlore deloyment challenges. We will also describe the detailed steps for preparing geofeeds for analysis, guiding future implementation. A key next step is using multiple geofeed snapshots over extended periods to study geofeed dynamics. Simplifying authentication requirements would help consumers leverage existing infrastructure, while providing publishers more flexibility to meet standards. Incorporating multiple secondary sources of prefix ownership will further refine and enhance our approach to geofeed authentication. 

\section*{Acknowledgments}
We would like to express our sincere gratitude to Professor Ramesh Govindan for his invaluable guidance and support throughout this research. His insights and expertise were crucial in shaping the direction of this work. We also wish to thank the members of the Program Committee for their valuable feedback and constructive comments, which greatly improved the quality of this paper.

\bibliographystyle{ieeetr}
\bibliography{noms2025}

\vspace{1cm}

\appendix
\subsection{Ethics}
This work deals with only publicly available data and does not raise any ethical issues.

\subsection{Data Visualization} \label{sec:analysis/viz}
To evaluate the gathered data, we visualize geofeed in two ways: a world map and a heatmap.
The first way is to have a world atlas that shows color gradient in a country-level, based on the IP prefix counts from \texttt{inetnum} and \texttt{inet6num}.
The second way is to have a plot where the x-axis is country and the y-axis is IP prefix from \texttt{inetnum} and \texttt{inet6num}. The original plots can be found in our \href{https://github.com/sulyabtv/geofeed-project/tree/main/graph}{GitHub}. The following are the four research questions that we come up with.

\subsubsection{World Map: do developed countries report geofeed more than developing countries?}
Our hypothesis in the World Map is that the published geofeed is predominantly from developed countries \cite{worldpopulationreviewDevelopedCountries, worldpopulationreviewDevelopingCountries}

We use \texttt{inetnum} and \texttt{inet6num} records from all five RIRs, namely AFRINIC, APNIC, ARIN, LACNIC and RIPE NCC.
For the World Map, we also use GIS (Geographic Information System) data from \href{https://hub.arcgis.com/datasets/esri::world-countries-generalized}{ARCGIS.com} to plot in a world map format.
We join those datasets together on ISO 3166 country codes.

\begin{figure*}[htbp]
  \centering
  \includegraphics[width=\linewidth,clip]{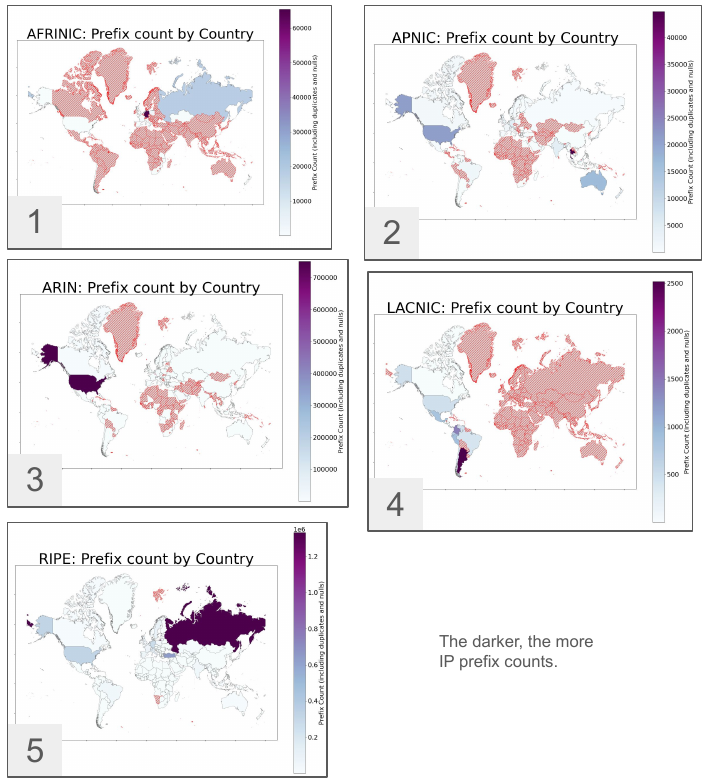}
  \caption{World Map by Prefix Count.\\\hspace{\textwidth}The country with the most IP prefix count in each RIR is 1: Germany in AFRNIC, 2: Thailand in APNIC, 3: US in ARIN, 4: Argentina in LACNIC, and 5: Russia in RIPE.}
  \label{fig:world_map}
\end{figure*}

As seen from Figure \ref{fig:world_map}, our hypothesis is not valid.
The published geofeed are mostly from developed countries, but not necessarily; they are darker in the color gradient.
The country with the most IP prefix count in each RIR is Germany in AFRNIC, Thailand in APNIC, US in ARIN, Argentina in LACNIC, and Russia in RIPE.
Most of the empty countries are from undeveloped countries; they are hatched with red lines.

It is interesting that the only a few countries from Africa have reported geofeed, even in AFRINIC.
Developed countries outside of Africa such as Germany, Russia, United States, and some European countries reported more than any country from Africa.
This discovery still validate our hypothesis, yet this can be a further research topic.

\subsubsection{Heatmap}
We use heatmap to understand geofeed in terms of the two IP address systems: IPv4 and IPv6.

As a methdology, we use \texttt{inetnum} and \texttt{inet6num} records from all five RIRs, namely AFRINIC, APNIC, ARIN, LACNIC and RIPE NCC.
We split them into IPv4 and IPv6.
We count the number of IP prefix by prefix length.
Particularly for IPv6, we only show prefix lengths that are multiples of 4.
Due to the skewed distribution over countries, we only show countries with at least 5\% portion of the overall IP prefix count in the relevant RIR.

\subsubsection{IPv4: is the commonly used IPv4 prefix /24 also common in geofeed?}
Our hypothesis is that /24 is the most common IPv4 prefix in geofeed.
/24 is indeed the most commonly used IPv4 prefix, because this provides a balance between the number of available hosts and efficient use of IP address space, making it suitable for small to medium-sized networks.

\begin{figure*}[htbp]
  \centering
  \includegraphics[width=\linewidth,clip]{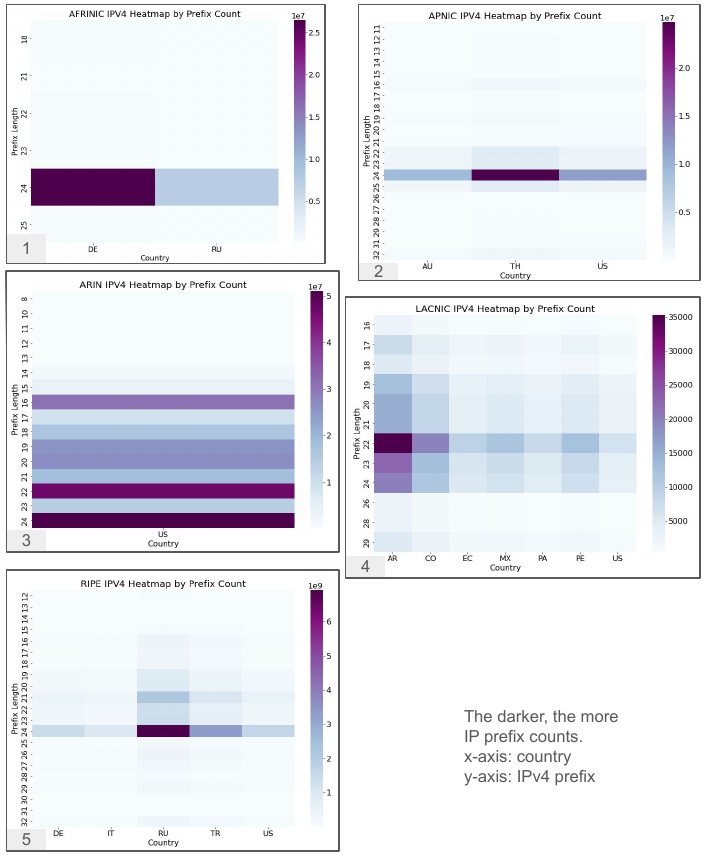}
  \caption{IPv4 Heatmap by Prefix Count.\\\hspace{\textwidth}The country with the most IP prefix count in each RIR is 1: Germany in /24 in AFRNIC, 2: Thailand in /24 in APNIC, 3: US in /24 in ARIN, 4: Argentina in /22 in LACNIC, and 5: Russia in /24 in RIPE.}
  \label{fig:ipv4_heatmap}
\end{figure*}

As seen from Figure \ref{fig:ipv4_heatmap}, our hypothesis for IPv4 is valid.
All RIRs but LACNIC show that /24 has the most number of IP prefix count.
The country with the most IP prefix count in each RIR is Germany in /24 in AFRNIC, Thailand in /24 in APNIC, US in /24 in ARIN, Argentina in /22 in LACNIC, and Russia in /24 in RIPE.

\subsubsection{IPv6: is the recommended IPv6 prefix /64 common in geofeed?}
Our hypothesis is that /64 is the most common IPv6 prefix in geofeed, because IETF (Internet Engineering Task Force) recommends that by being selective among prefixes /48, /64, and /128 from a previously announced RFC~\cite{narten2011rfc}.

\begin{figure*}[htbp]
  \centering
  \includegraphics[width=0.99\textwidth,clip]{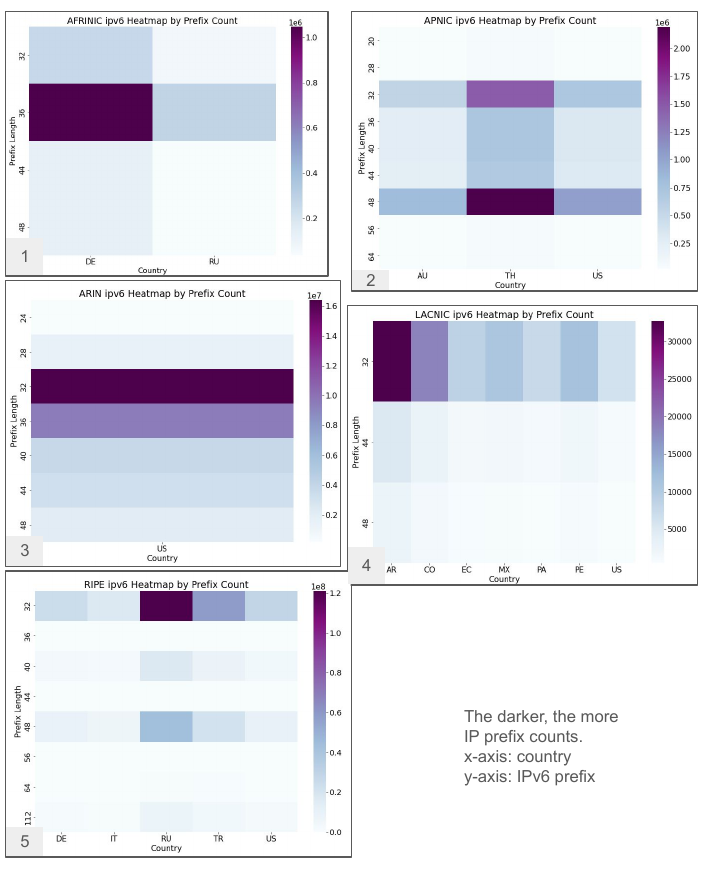}
  \caption{IPv6 Heatmap by Prefix Count. \\\hspace{\textwidth}The country with the most IP prefix count in each RIR is 1: Germany in /36 in AFRNIC, 2: Thailand in /48 in APNIC, 3: US in /32 in ARIN, 4: Argentina in /32 in LACNIC, and 5: Russia in /32 in RIPE.}
  \label{fig:ipv6_heatmap}
\end{figure*}

As seen from Figure \ref{fig:ipv6_heatmap}, our hypothesis for IPv6 is not valid.
It turns out that /32 is the most commonly reported prefix in IPv6.
The country with the most IP prefix count in each RIR is Germany in /36 in AFRNIC, Thailand in /48 in APNIC, US in /32 in ARIN, Argentina in /32 in LACNIC, and Russia in /32 in RIPE.

\end{document}